# Dual Watermarking Scheme with Encryption

R.Dhanalakshmi
PG Scholar
Dept of CSE
Sri Venkateswara college
Of Engineering
Post Bag No. 3, Pennalur
Sriperumbudur
602 105
India

K.Thaiyalnayaki
Assistant Professor
Dept of IT
Sri Venkateswara college
of Engineering
Post Bag No. 3, Pennalur
Sriperumbudur
602 105
India

*Abstract-* *Digital Watermarking is used for copyright protection and authentication. In the proposed system, a Dual Watermarking Scheme based on DWT-SVD with chaos encryption algorithm, will be developed to improve the robustness and protection along with security. DWT and SVD have been used as a mathematical tool to embed watermark in the image. Two watermarks are embedded in the host image. The secondary is embedded into primary watermark and the resultant watermarked image is encrypted using chaos based logistic map. This provides an efficient and secure way for image encryption and transmission. The watermarked image is decrypted and a reliable watermark extraction scheme is developed for the extraction of the primary as well as secondary watermark from the distorted image.*

## I. INTRODUCTION

The process of embedding information into another object/signal can be termed as watermarking. Watermarking is mainly used for copy protection and copyright-protection. Historically, watermarking has been used to send ``sensitive" information hidden in another signal. Watermarking has its applications in image/video copyright protection. The characteristics of a watermarking algorithm are normally tied to the application it was designed for. The following explain the requirements of watermarking: i) Imperceptibility - A watermark is called perceptible if its presence in the marked signal is noticeable, but non-intrusive. A watermark is called imperceptible if the cover signal and marked signal are indistinguishable with respect to an appropriate perceptual metric.
ii) Robustness - The watermark should be able to survive any reasonable processing inflicted on the carrier. A watermark is called fragile if it fails to be detected after the slightest modification. Fragile watermarks are commonly used for tamper detection (integrity proof).
iii) Security - The watermarked image should not reveal any clues of the presence of the watermark, with respect to un-authorized detection, or undetectability or unsuspicious.

A visible watermark howsoever robust it may be can always be tampered using various software. To detect such kind of tampering (in worst case to protect the image when the visible watermark is fully removed) an invisible watermark can be used as a back up. The dual watermark is a combination of a visible watermark and an invisible watermark. The visible watermark is first inserted in the original image and then an invisible watermark is added to the already visible-watermarked image. The final watermarked image is the dual watermarked image.

The first applications were related to copyright protection of digital media. In the past duplicating artwork was quite complicated and required a high level of expertise for the counterfeit to look like the original. However, in the digital world this is not true. Now it is possible for almost anyone to duplicate or manipulate digital data and not lose data quality. Similar to the process when artists creatively signed their paintings with a brush to claim copyrights, artists of today can watermark their work by hiding their name within the image. Hence, the embedded watermark permits identification of the owner of the work.

With the growing threat of piracy in the Internet and copyright infringement cases, watermarks are sure to serve an important role in the future of intellectual property rights.

## II. PROPOSED SYSTEM

Dual watermarking scheme based on DWT and Singular Value Decomposition (SVD) along with the chaos based encryption technique is proposed. After decomposing the cover image into four bands (LL, HL, LH, and HH), we apply the SVD to each band, and modify the singular values of the cover image with the singular values of the watermarked primary watermark. When the primary watermark image is in question, the invisible secondary watermark can provide rightful ownership. Modification in all frequencies allows the development of a watermarking scheme that is robust to a wide range of attacks. SVD transform is performed on all the images and sum up the singular values to find the new singular values. Both the watermarks are embedded in the same manner and the watermarked primary watermark is encrypted using chaos encryption.

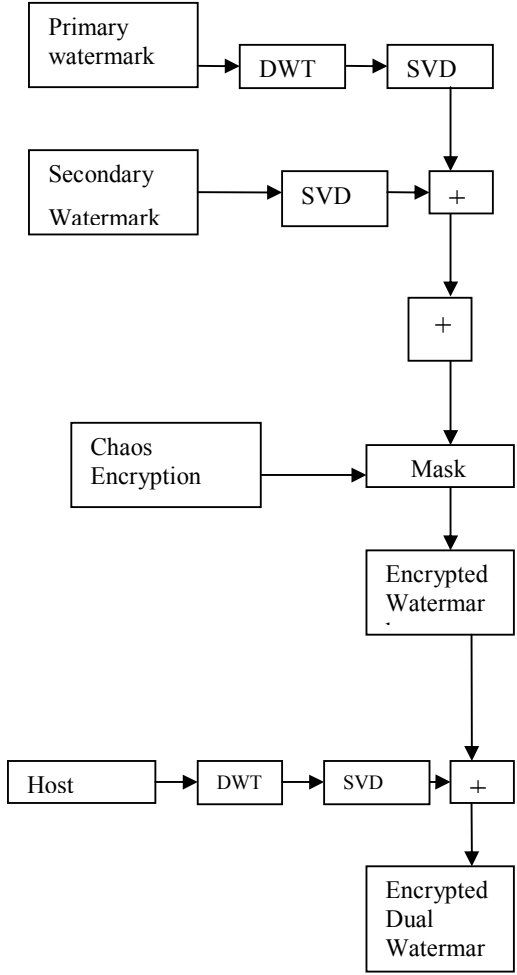

Fig. 1 Block Diagram of the proposed system

### III. MODULES

In the proposed system, there are four modules, they are as follows:
1. Embedding secondary watermark into primary.
2. Encryption of watermarked primary image and embedding in the host image.
3. Attacks
4. Extraction of primary and secondary watermark from the host image.

*A. Embedding secondary watermark into primary watermark*

In two-dimensional DWT, each level of decomposition produces four bands of data denoted by LL, HL, LH, and HH. The LL subband can further be decomposed to obtain another level of decomposition. This process is continued until the desired number of levels determined by the application is reached. In DWT-SVD based watermarking, the singular values of the detail and approximate coefficients are extracted. The extracted singular values are modified to embed the watermark data.

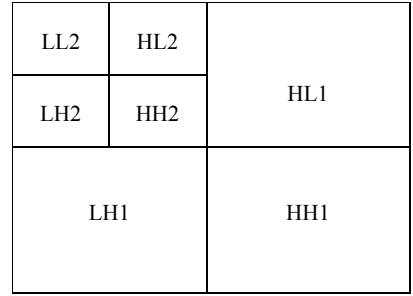

Fig. 2 DWT

Let A be a general real matrix of order m × n. The singular value decomposition (SVD) of A is the factorization:

$$A = U * S * V^T \quad (1)$$

where U and V are orthogonal(unitary) and $S = diag(\sigma_1, \sigma_2, ..., \sigma_r)$, where $\sigma_i$, $i = 1(1)r$ are the singular values of the matrix A with $r = min(m, n)$ and satisfying :

$$\sigma_1 \geq \sigma_2 \geq ... \geq \sigma_r \quad (2)$$

The first r columns of V the right singular vectors and the first r columns of U the left singular vectors.

Use of SVD in digital image processing has some advantages. First, the size of the matrices from SVD transformation is not fixed. It can be a square or a rectangle. Secondly, singular values in a digital image are less affected if general image processing is performed. Finally, singular values contain intrinsic algebraic image properties.

The singular values are resistant to the following types of geometric distortions:
i) Transpose: The singular value matrix A and its transpose $A^T$ have the same non-zero singular values.
ii) Flip: A, row-flipped $A_{rf}$, and column-flipped $A_{cf}$ have the same non-zero singular values.
iii) Rotation: A and $A_r$ (A rotated by an arbitrary degree) have the same non-zero singular values.
iv) Scaling: B is a row-scaled version of A by repeating every row for $L_1$ times. For each non-zero singular value λ of A, B has $L_1\lambda$ . C is a column-scaled version of A by repeating every column for $L_2$ times. For each nonzero singular value λ of A, C has $L_2 \lambda$. If D is row-scaled by $L_1$ times and column-scaled by $L_2$ times, for each non-zero singular value λ of A, D has L1L2λ.

v) Translation: A is expanded by adding rows and columns of black pixels. The resulting matrix $A_e$ has the same non-zero singular values as A.

*Algorithm – Embedding watermark.*

i) Perform 1-level wavelet transform on the primary watermark.
Let us denote each sub-band with $W_{2\theta}$ where $\theta \varepsilon$ {LL, LH, HL, HH} represents the orientation.
ii) Perform SVD transform on secondary watermark,

$$W2 = U_{W2} * S_{W2} * V^T_{W2}.$$

iii) Perform SVD transform on approximation and all the detail parts of the primary watermark,

$$W1_\theta = U_{W1\theta} * S_{W1\theta} * V^T_{W1\theta}$$

where $\theta \varepsilon$ { LL, LH, HL, HH }.
iv) Modify the singular values of approximation and all the detail parts with the singular values of the secondary watermark as

$$S^*_\theta = \alpha S_{W2} + S_{W1\theta}$$

v) Obtain modified approximation and all the detail parts as

$$W^*_{1\theta} = U_{W1\theta} * S^*_{W1\theta} * V^T_{W1\theta}$$

where $\theta \varepsilon$ { LL, LH, HL, HH }.
vi) Perform 1-level inverse discrete wavelet transform to get the watermarked primary watermark.

*B. Encryption*

Chaos theory is a branch of mathematics which studies the behavior of certain dynamical systems that may be highly sensitive to initial conditions. This sensitivity is popularly referred to as the butterfly effect. As a result of this sensitivity, which manifests itself as an exponential growth of error, the behavior of chaotic systems appears to be random. That is, tiny differences in the starting state of the system can lead to enormous differences in the final state of the system even over fairly small timescales. This gives the impression that the system is behaving randomly.

Chaos-based image encryption techniques are very useful for protecting the contents of digital images and videos. They use traditional block cipher principles known as chaos confusion, pixel diffusion and number of rounds. The complex structure of the traditional block ciphers makes them unsuitable for real-time encryption of digital images and videos. Real-time applications require fast algorithms with acceptable security strengths. The chaotic maps have many fundamental properties such as ergodicity, mixing property and sensitivity to initial condition/system parameter and which can be considered analogous to some cryptographic properties of ideal ciphers such as confusion, diffusion, balance property.
A chaos-based image encryption system based on logistic map, in the framework of stream cipher architecture, is proposed. This provides an efficient and secure way for image encryption and transmission.

*Logistic Map*

One of the chaos functions that have been studied recently for cryptography applications is the logistic map. The logistic map function is expressed as:

$$X_{n+1} = rXn (1-X_n)$$

where $X_n$ takes values in the interval [0,1]. It is one of the models that present chaotic behavior. The parameter *r* can be divided into three segments.
When $X_0$=0.3 and r ε [0,3] the system works normal without any chaotic behavior. When r ε [3, 3.57], the system appears periodicity. While r ε [3.57, 4]*,* it becomes a chaotic system with no periodicity.
We can draw the following conclusions:
i) When r ε [0, 3.57], the points concentrate on several values and could not be used for image cryptosystem.
ii) For r ε [3.57, 4], the logistic map exhibits chaotic behavior, and hence the property of sensitive dependence. So it can be used for image cryptosystem.

*Algorithm – Image Encryption*

i) The watermarked primary image is converted to a binary data stream.
ii) A random keystream is generated by the chaos-based pseudo-random keystream generator (PRKG).
iii) PRKG is governed by a logistic map, which is depended on the values of b, $x_0$.
iv) Through iterations, the first logistic map generates a hash value $x_{i+1}$, which is highly dependent on the input (b, $x_0$), is obtained and used to determine the system parameters of the second logistic map.
v) The real number $x_{i+1}$ is converted to its binary representation $x_{i+1}$, suppose that L=16, thus $x_{i+1}$ is {b1, b2, b3, … b16}. By defining three variables whose binary representation is $X_l$=b1…b8, $X_h$=b9…b16, we obtain $X_{i+1}'=X_l \oplus X_h$.
vi) Mask the watermarked primary image with the chaos values.
The generator system can be briefly expressed using the following logistic maps:

$x_{i+1}=bx_i(1-x_i)$                             (1)

$x_{i+1}' = X_{i+1}' = X_l \oplus X_h$                (2)

$WI'=WI_i \oplus X_{i+1}'$                         (3)

The encrypted watermarked primary image is then embedded into the host image and transmitted.

*C. Attacks*

To investigate the robustness of the algorithm, the watermarked image is attacked by Average and Mean Filtering, JPEG and JPEG2000 compression, Gaussian noise addition, Resize, Rotation and Cropping. After these attacks on the watermarked image, we compare the extracted watermarks with the original one. The watermarked image quality is measured using PSNR (Peak Signal to Noise Ratio).

*D. Extraction of watermarks*

Decryption is the reverse iteration of encryption. After decryption of the watermarked primary image, the extraction process takes place.

*Extracting Primary Watermark*

The extraction technique for primary watermark is given asvfollows:

i)Perform 1-level wavelet transform on the host and the watermarked image. Denote each sub-band with $W_\theta$ and $\hat{W}_\theta$ for host and watermarked image respectively where $\theta \in \{LL, LH, HL, HH\}$ represents the orientation.

ii) The detail and approximation sub-images of the host as well as watermarked image is segmented into non overlapping rectangles.

iii) Perform SVD transform on all non overlapping rectangles of both images.

$$W_{1\theta} = U_{W\theta} * S_{W\theta} * V^T_{W\theta} \text{ and}$$

$$\hat{W}_\theta = U_{\hat{w}\theta} * S_{\hat{w}\theta} * V^T_{\hat{w}\theta}$$

where $\theta \in \{LL, LH, HL, HH\}$.

iv)Extract singular values from primary watermark from all non overlapping rectangles as

$$S = S_{\hat{w}\theta} - S_{W\theta}/\beta$$

v) The primary watermark can be obtained as

$$W1' = U_{W1} * S * V^T_{W1}$$

*Extracting Secondary Watermark*

i)Perform 1-level wavelet transform on the primary watermark and the detected watermark W1'. Denote each sub-band with $W'_{1\theta}$ and $W_{1\theta}$ where $\theta \in \{LL, LH, HL, HH\}$ represents the orientation.

ii)Perform SVD transform on all subbands.

iii)The singular values of secondary watermark can be extracted as

$$S' = S' W'_{1\theta} - SW_{1\theta}/\alpha$$

iv) The secondary watermark can be obtained as

$$W2 = U_{W2} * S' * V^T_{W2}$$

IV. RESULTS AND DISCUSSION

The proposed algorithm is demonstrated using MATLAB. We have taken 8-bit gray scale tree image as host image of size 256 x 256 and for primary and
secondary watermark, we have used 8-bit gray scale lena image and boy image of sizes 128 × 128 and 64 × 64 respectively. The secondary watermark is embedded into primary and the watermarked primary is encrypted. For encryption, chaos encryption technique is used.

For embedding the encrypted watermarked primary into the host image, we have used 2-level of decomposition using Daubechies filter bank. For extracting both the watermarks, decryption is done using the chaos technique. The decrypted image is then used to extract the primary watermark and this is used for extracting the secondary watermark. In figures 2 and 3 all original, watermarked images and extracted watermarks are shown.

To investigate the robustness of the algorithm, the watermarked image is attacked by Average and Median Filtering, Gaussian noise addition, Resize and Rotation. After these attacks on the watermarked image, we compare the extracted watermarks with the original one.

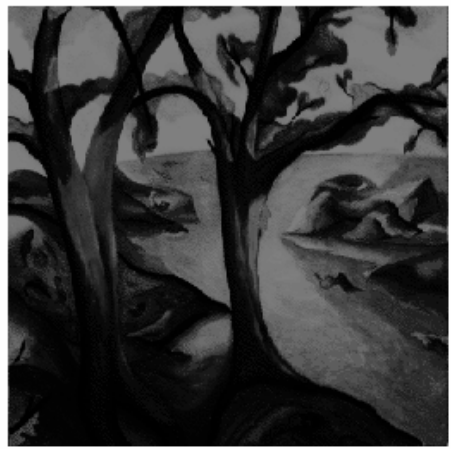

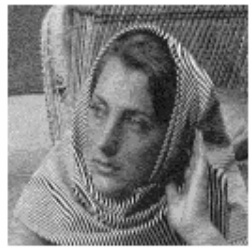 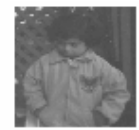

Fig. 2. Original Images a) Host image b) Primary watermark c) Secondary watermark

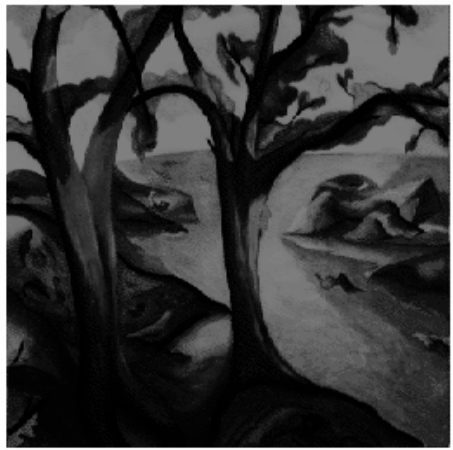

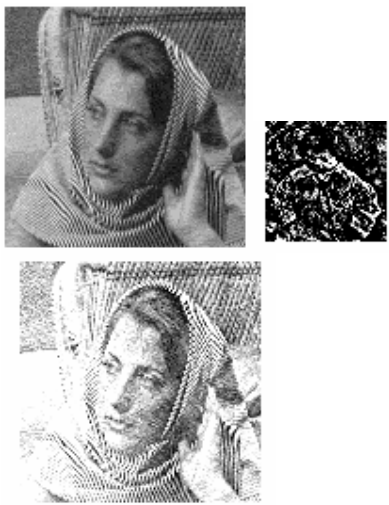

Fig. 3. Watermarked and extracted watermark images a) Watermarked Host image b) Watermarked Primary watermark c) Extracted Secondary watermark d) Extracted Primary watermark

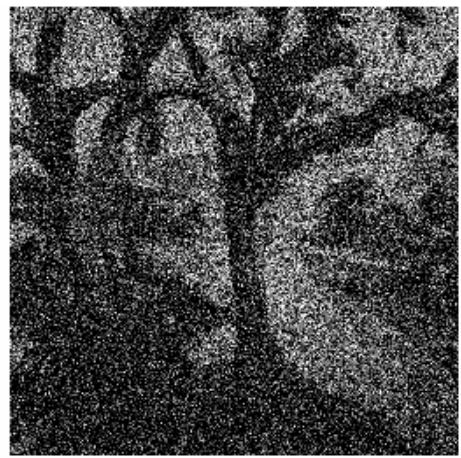

Fig. 5. Additive Gaussian Noise Attack a) Attacked Host image b) Extracted Primary watermark c) Extracted Secondary watermark

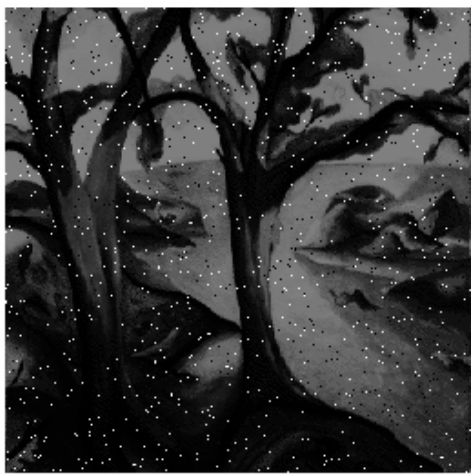

Fig. 4. Median filtering Attack a) Attacked Host image b) Extracted Primary watermark c) Extracted Secondary watermark

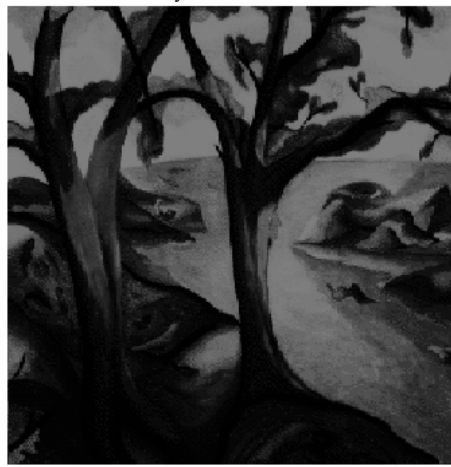

Fig. 6. Resize Attack (512 X 512) a) Attacked Host image b) Extracted Primary watermark c) Extracted Secondary watermark

on coupled chaotic logistic maps that one logistic chaotic system generates the random changing parameter to control the parameter of the other. The watermarked primary image is encrypted using the chaos based encryption technique. Later it is embedded in the cover image and transmitted. The chaotic encryption scheme supplies us with a wide key space, high key sensitivity, and the cipher can resist brute force attack and statistical analysis. It is safe and can meet the need of image encryption.

For the extraction of watermark, a reliable watermark decryption scheme and an extraction scheme is constructed for both primary and secondary watermark. Robustness of this method is carried out by variety of attacks.

## V. REFERENCES

[1] Gaurav Bhatnagar, Balasubramanian Raman and K. Swaminathan, " DWT-SVD based Dual Watermarking Scheme", IEEE International Conference on the Applications of Digital Information and Web Technologies (ICADIWT-2008), pp. 526-531.

[2] R. Liu and T. Tan, "An SVD-Based Watermarking Scheme for Protecting Rightful Ownership," IEEE Transactions on Multimedia, vol. 4, no. 1, 2002, pp. 121-128.

[3] E. Ganic and A. M. Eskicioglu, "Robust Embedding of Visual Watermarks Using DWT-SVD," Journal of Electronic Imaging, vol. 14, no. 4, 2005.

[4] Shubo Liu, Jing Sun, Zhengquan Xu, Jin Liu, "Analysis on an Image Encryption Algorithm", 2008 International Workshop on Education Technology and Training & 2008 International Workshop on Geoscience and Remote Sensing, pp. 803- 806.

[5] Hossam El-din H. Ahmed, Hamdy M. Kalash, and Osama S. Farag Allah, "An Efficient Chaos-Based Feedback Stream Cipher (ECBFSC) for Image Encryption and Decryption", Informatica (2007) 121–129.

[6] Xiao-jun Tong, Ming-gen Cui, "A New Chaos Encryption Algorithm Based on Parameter Randomly Changing", IFIP International Conference on Network and Parallel Computing ( 2007) 303-307.

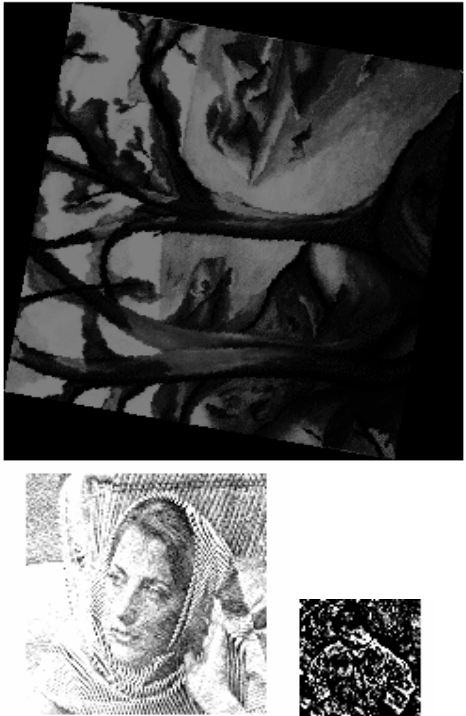

Fig. 7. Rotation Attack (80° Rotation) a) Attacked Host image b) Extracted Primary watermark c) Extracted Secondary watermark

Table 1. Correlation Coefficient of Extracted Primary and Secondary Watermark

| Attacks | Primary | Secondary |
|---|---|---|
| Median Filtering | 0.8967 | 0.4157 |
| Additive Gaussian | 0.8966 | 0.4161 |
| Resize | 0.8968 | 0.4154 |
| Rotation | 0.8969 | 0.4153 |

## V. CONCLUSION

This paper deals with a novel dual watermarking scheme, which includes encryption, to improve rightful ownership, protection and robustness. An image encryption algorithm based on logistic map is proposed. A well-designed chaos-based stream cipher can be a good candidate and may even outperform the block cipher, on speed and security. In this, the key stream generator is based